\documentclass[aps,twocolumn,superscriptaddress,preprintnumbers,showpacs,article,prl,longbibliography]{revtex4-1}

\usepackage[latin9]{inputenc}
\usepackage{color}
\usepackage{verbatim}
\usepackage{amsmath}
\usepackage{amssymb}
\usepackage{graphicx}
\usepackage{esint}
\usepackage{esvect}
\usepackage{braket}
\usepackage{amsmath,bm}
\usepackage{graphicx}
\usepackage{subfigure}
\usepackage{url}
\usepackage{lipsum}
\usepackage{lipsum}
\usepackage[dvipsnames]{xcolor}
\usepackage{hyperref}
\usepackage[resetlabels]{multibib}

\makeatletter
\@ifundefined{textcolor}{}
{%
 \definecolor{BLACK}{gray}{0}
 \definecolor{WHITE}{gray}{1}
 \definecolor{RED}{rgb}{1,0,0}
 \definecolor{GREEN}{rgb}{0,1,0}
 \definecolor{BLUE}{rgb}{0,0,1}
 \definecolor{CYAN}{cmyk}{1,0,0,0}
 \definecolor{MAGENTA}{cmyk}{0,1,0,0}
 \definecolor{YELLOW}{cmyk}{0,0,1,0}
}

\usepackage{mathrsfs}

\draft
\makeatother
\begin{document}

\title{Breakdown of helical edge state topologically protected conductance in time-reversal-breaking excitonic insulators}
\author{Yan-Qi Wang}
\affiliation{Department of Physics, University of California,  Berkeley, California 94720, USA}
\affiliation{Materials Sciences Division, Lawrence Berkeley National Laboratory, Berkeley, California 94720, USA}
\author{Micha\l{} Papaj}
\affiliation{Department of Physics, University of California,  Berkeley, California 94720, USA}
\author{Joel E. Moore}
\affiliation{Department of Physics, University of California,  Berkeley, California 94720, USA}
\affiliation{Materials Sciences Division, Lawrence Berkeley National Laboratory, Berkeley, California 94720, USA}

\begin{abstract}
Gapless helical edge modes are a hallmark of the quantum spin Hall effect.  Protected by time-reversal symmetry, each edge contributes a quantized zero-temperature conductance quantum $G_0 \equiv e^2/h$. However, the experimentally observed conductance in WTe$_2$ decreases below $G_0$ per edge already at edge lengths around 100 nm, even in the absence of explicit time-reversal breaking due to an external field or magnetic impurities. In this work, we show how a time-reversal breaking excitonic condensate with a spin-spiral order that can form in WTe$_2$ leads to the breakdown of conductance quantization. We perform Hartree-Fock calculations to compare time-reversal breaking and preserving excitonic insulators. Using these mean-field models we demonstrate via quantum transport simulations that weak non-magnetic disorder reproduces the edge length scaling of resistance observed in the experiments. We complement this by analysis in the Luttinger liquid picture, shedding additional light on the mechanism behind the quantization breakdown.
\end{abstract}

\maketitle

\section{Introduction}
The discovery of the quantum spin Hall (QSH) effect gave rise to a flurry of research on the topological aspects of quantum materials behavior. One of the key features of the QSH is the presence of helical edge states protected by the combination of symmetry and topology~\cite{Kane2005A,Kane2005B,Bernevig2006,Hasan2010,Qi2011,Alicea2012,Ren2016,Culer2020,Vayrynen2018}. The low-energy spectrum of such a system consists of counterpropagating electron states with opposite spins, connected by time-reversal symmetry.  Due to the orthogonality of states in a Kramers pair, elastic backscattering by a static potential is forbidden as long as time-reversal symmetry is preserved. When the bulk of the material is insulating, each such helical edge state yields a quantized zero-temperature conductance given by $G_0 \equiv e^2/h$, the conductance quantum. The first theoretically predicted QSH insulators were HgTe/CdTe quantum well heterostructures, which have been intensively studied experimentally since the initial proposal~\cite{Konig2007quantum,Roth2009,Knex2011,Spanton2014,Pribiag2015edge,Du2015,Li2015,Du2017,Du2017evidence,Bendias2018high,Lunczer2019,Xiao2019,Han2019,Piatrusha2019,Strunz2020interacting,Shamim2020emergent,Dartiailh2020,Shamim2021quantized}. However, because the penetration depth of the edge states in HgTe quantum wells is large, it is often difficult to separate the edge physics from the bulk behavior~\cite{bieniek2022theory}. This has lead to the search for alternative platforms, among which the most prominent are single layers of the transition metal dichalcogenide WTe$_2$ in the 1T' crystalline structure \cite{Qian2014,Fei2017,Tang2017,Wu2018,Li2020}. Several experiments have observed signatures of the QSH effect in that platform, both through transport measurements~\cite{Fei2017,Wu2018} as well as scanning tunneling microscopy/spectroscopy~\cite{Tang2017,Jia2017,peng2017observation}.

Unfortunately, while the edge transport provides a new way towards dissipationless transport and quantum computation~\cite{Kane2005A,Kane2005B,Bernevig2006,Hasan2010,Qi2011,Alicea2012,Ren2016,Culer2020,Vayrynen2018}, the experimental observation of robust conductance quantization in realistic scenarios has been elusive~\cite{Markus2007,Roth2009,Gusev2014,Knez2014,Du2015,Nichele2016,Du2017,Fei2017,Wu2018}. For example, in the case of WTe$_2$, while signs of QSH have been observed up to temperatures of 100 K, these pertain to the devices with edge lengths shorter than 100 nm, much smaller than the multiple-micron lengths for the conductance quantization seen in the integer quantum Hall effect, which also is topological in origin. This discrepancy between the expected robustness of quantization and the imperfect experimental behavior prompted intense theoretical exploration of the possible explanations for this difference. One of the fundamental reasons for the deviation from perfect quantization is that while the time-reversal symmetry precludes the overlap of wave functions of counter-propagating degenerate time-reversal states, there is no such restriction for them at different energies. Therefore, time-reversal-invariant perturbations can still lead to back-scattering of the electrons in a helical channel through interaction-induced inelastic processes with the help of non-magnetic disorder~\cite{Hsu_2021}. In fact, interaction-induced inelastic one- or two-particle backscattering is allowed since the momentum difference between the initial and final states can be compensated by non-magnetic disorder~\cite{Schmidt2012,Lezmy2012}. The deviation of the perfect conductance quanta has been thus attributed to many factors, including coupling to charge puddles~\cite{Vayrynen2013,Vayrynen2014}, incoherent electromagnetic noise~\cite{Vayrynen2018}, nuclear spins~\cite{Hsu2017,Hsu2018}, quenched disorder~\cite{Wu2006,Xu2006}, spin orbit coupling~\cite{Strom2010,Geissler2014,Kainaris2014,Xie2016,Kharitonov2017,DelMaestro2013,Crepin2012}, and spin-phonon coupling~\cite{Budich2012,Groenendijk2018}.

In this work we explore the possibility of explaining the deviation from the perfect conductance quantization in WTe$_2$ via the formation of a time-reversal breaking excitonic condensate \cite{blasonExcitonTopologyCondensation2020, Kwan2021}. Besides the quantum spin Hall effect, WTe$_2$ exhibits also fascinating interaction-driven effects, including superconductivity \cite{fatemiElectricallyTunableLowdensity2018} and potential excitonic insulator states \cite{sunEvidenceEquilibriumExciton2022, jiaEvidenceMonolayerExcitonic2022}. In the latter case, the effect is due to the possible semimetallic noninteracting band structure of WTe$_2$ with a hole pocket around the $\Gamma$ point of the Brillouin zone and two electron pockets along the $\Gamma-X$ direction. Formation of an excitonic condensate with finite momentum pairing equal to the separation between the pockets was postulated and experimental signatures of such a state were observed \cite{sunEvidenceEquilibriumExciton2022, jiaEvidenceMonolayerExcitonic2022}.

However, the exact nature of the excitonic state is unclear and the possibility of time-reversal breaking spin-spiral or spin-density wave at the Hartree-Fock mean field level has been raised \cite{Kwan2021}.  Starting from a bulk Hartree-Fock calculation, we observe both time-reversal-breaking and -preserving energy minima, with unconstrained minimization often favoring the former.  We then derive a tight-binding model for the excitonic insulator states and use it to perform quantum transport calculations for a finite width ribbon with disorder. We demonstrate that while the time-reversal-preserving excitonic insulator is topological and thus exhibits robust conductance quantization of edge state transport, the time-reversal-breaking condensate deviates from $e^2/h$ per edge state conductance in the presence of non-magnetic static disorder. However, the remnants of the helical edge states, though unprotected from backscattering, remain in the exciton-induced gap and allow us to reproduce the experimentally observed edge-length scaling of resistance, with results close to quantized below 100 nm but with the deviation increasing substantially for longer edges. We then supplement these simulations by analysis in the Luttinger liquid picture, shedding additional light on the mechanisms that lead to the breakdown of conductance quantization.

\section{Transport from the bulk theory}\label{SecBulkTransport}

To perform the quantum transport simulations for the excitonic insulator phases of WTe$_2$, we employ a tight-binding model that is the finite difference approximation of a continuum model given by:
\begin{align}
H_0(\mathbf{k}) = &\left(a k_x^2 + b k_x^4 + 2 b k_x^2 k_y^2 + b_y k_y^4 + \frac{\delta}{2}\right) I_d \notag \\ &+ \left(-\frac{\mathbf{k}^2}{2m} - \frac{\delta}{2}\right) I_p + v_x k_x \tau_x s_y + v_y k_y \tau_y s_0 
\end{align}
where $\mathbf{k}^2 = k_x^2 +k_y^2$, $\tau_i$ and $s_i$ are Pauli matrices in $p$, $d$ orbitals and spin spaces, respectively, $I_d = (\tau_0 + \tau_z)/2\, s_0$ and $I_p = (\tau_0 - \tau_z)/2\, s_0$ are identity matrices for $d$ and $p$ orbitals, while $v_x$ and $v_y$ determine the spin-orbital coupling. The parameter values we use in the calculations are $a=-3, b=18, b_y=40, \delta=-0.9, m=0.03, v_x=0.5, v_y=3$, where all the energies are expressed in eV and lengths in {\AA}. The different value of $b_y$ parameter as compared to Ref.~\cite{jiaEvidenceMonolayerExcitonic2022} was chosen to ensure that the low energy behavior of the full lattice model is consistent with the continuum model within the cutoff employed therein, with no extra low-energy valley along the $\Gamma-Y$ direction. 

We discretize this Hamiltonian on a rectangular lattice with lattice constants $a_x = 2.805\,${\AA} and $a_y = 6.27\,${\AA}. In discretizing the Hamiltonian we use the finite difference terms up to ($\pm 3, \pm 3$) hoppings in $x$ and $y$ directions, respectively. At charge neutrality point this system has a hole pocket around $\Gamma$ point and two electron pockets with minima at $\mathbf{q}_c=\pm 0.32\, \hat{x}$ along the $\Gamma-X$ direction of the Brillouin zone. Based on previous works \cite{jiaEvidenceMonolayerExcitonic2022, Kwan2021}, we expect the formation of an excitonic condensate at finite momentum corresponding to the pocket separation $q_c$ in momentum space. When time-reversal symmetry is not enforced, the excitonic order can form either a spin spiral or spin density wave phases, depending on the interaction strength. The lattice constant $a_x$ was therefore chosen such that for $q_c$ as determined for WTe$_2$ from first principles the resulting order would be commensurate with the discretized lattice with a period increased by a factor of $L_x = 7$. This simplifies expressing the model with mean field order parameter in real space. When a finite $q$ order is allowed, the period of the lattice increases correspondingly and the Brillouin zone (BZ) shrinks, while the electron bands are folded into the smaller BZ. We can then label these bands by their corresponding momenta $\mathbf{k}$, spin and orbital index $\alpha$, and finally the reciprocal lattice vector of the enlarged unit cell $\mathbf{G}_i$, which indicates from which extended Brillouin zone the particular state comes from. In other words, the original momentum $\mathbf{k}_0$ of the state before folding becomes decomposed as $\mathbf{k}_0 = \mathbf{k} + \mathbf{G}_i$.

\begin{figure*}
    \centering
    \includegraphics[width=0.99\linewidth]{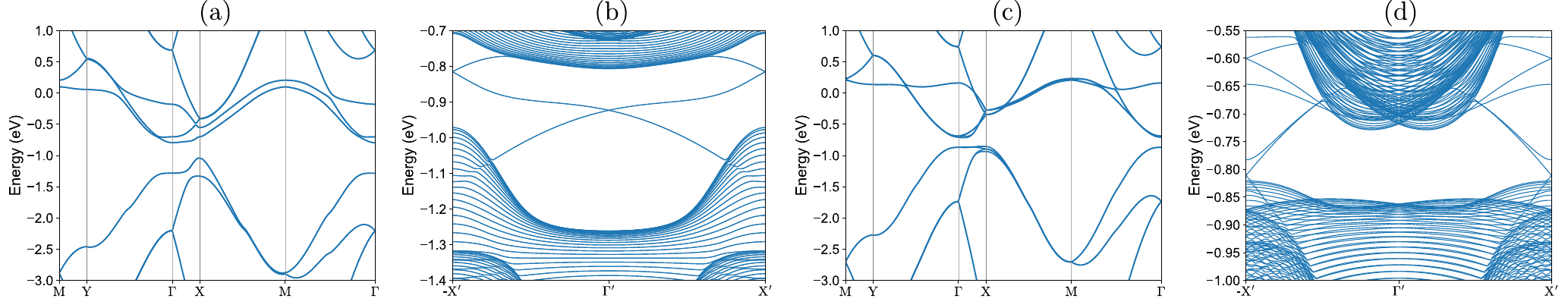}
    \caption{The band structures of the excitonic condensate systems within Hartree-Fock mean field approximation. (a) 2D tight-binding model with time-reversal symmetry preserved. The gap is opened by the exciton formation, but the bands remain doubly degenerate due to symmetry preservation. (b) Spectrum of system from (a) placed on a quasi-1D ribbon. The system is topological, which results in presence of helical edge states within the excitonic band gap. (c) 2D tight-binding model with time-reversal symmetry broken. The excitons open the gpa, but the spin degeneracy is lifted due to formation of the spin spiral state. (d) Spectrum of system from (c) placed on a quasi-1D ribbon. Even though time-reversal is broken, as the magnitude of symmetry breaking terms is relatively small, the remainder of the helical edge states remain in the gap.}
    \label{fig:bands}
\end{figure*}

We then consider Coulomb interaction of the electrons in the tight-binding model given by:
\begin{equation}
H_\mathrm{int} = \frac{1}{2N_{k_0}\Omega} \sum_{\mathbf{k}_0, \mathbf{p}_0, \mathbf{q}_0} \sum_{\alpha, \beta} V_{\mathbf{q}_0} c^\dagger_{\mathbf{k}_0+\mathbf{q}_0,\alpha} c^\dagger_{\mathbf{p}_0-\mathbf{q}_0,\beta} c_{\mathbf{p}_0,\beta} c_{\mathbf{k}_0, \alpha}
\end{equation}
where $\Omega$ is the area of the crystal unit cell, $N_{k_0}$ is the number of momentum points taken in the summation, $c^\dagger_{\mathbf{k}_0,\alpha}$ are creation operators of a particle with momentum $\mathbf{k}_0$ and $\alpha$ labeling both the orbitals and spin. We assume double-gate screening of the Coulomb potential, with the resulting Fourier transform:
\begin{equation}
V_\mathbf{q} = V_0 \frac{\tanh{\xi q / 2}}{\xi q / 2}
\end{equation}
with $\xi=250\,${\AA} being the distance between the gates and $V_0$ the interaction strength parameter. We then include the impact of Coulomb interaction at the mean field level by performing Hartree-Fock calculations. To remain consistent with the previous Hartree-Fock calculations for this model, even though we work with a tight-binding lattice model, we still maintain the cutoff in momentum summation as in continuum model of Refs.~\cite{jiaEvidenceMonolayerExcitonic2022, Kwan2021}, with $|k_x| < 3/2\, q_c$ and $|k_y|<0.25$. With the cutoff imposed, we also appropriately rescale the interaction strength parameter to reflect the decreased number of momentum points within the cutoff. We compare the results when we allow for breaking of inversion and time-reversal symmetries during the self-consistent calculation to the case where time-reversal is preserved. We choose the interaction strength parameter $V_0$ so that the rescaled interaction energy within the cutoff at wave vector $q_c$ is $\tilde{V}_{q_c} = 1.71$eV in the case with time-reversal symmetry enforced, and $\tilde{V}_{q_c} = 0.9\,$eV otherwise. In expressing the rescaled interaction strength, we follow the convention of Ref.~\cite{jiaEvidenceMonolayerExcitonic2022}. In both cases we consider the formation of finite momentum order parameter at $q_c = 0.32$ along $x$ direction as mentioned above. In each self-consistent iteration we diagonalize the quadratic Hamiltonian $H_\mathrm{MF} = H_0 + H_\mathrm{HF}$ to obtain a set of $n=4L_x$ eigenvalues $\epsilon_{\mathbf{k}n}$ at each momentum and their corresponding eigenvectors $u_{\mathbf{k}\mathbf{G}\alpha n}$, which are $n$ component spinors. These eigenvalues and eigenvectors are then used to obtain the order parameters according to:

\begin{align}
&\Delta_{\mathrm{H}\alpha\beta}^{\mathbf{G}_1\mathbf{G}_2} =\notag \\ &\delta_{\alpha\beta}\sum_{\mathbf{p}\mathbf{G}n}\frac{V(\mathbf{G}_1-\mathbf{G}_2)}{N_k} u^*_{\mathbf{p}\mathbf{G}-\mathbf{G}_1+\mathbf{G}_2 \beta n}u_{\mathbf{p}\mathbf{G}\beta n} f_0(\epsilon_{\mathbf{p}n})
\end{align}

\begin{align}
&\Delta_{\mathrm{F}\alpha\beta}^{\mathbf{G}_1\mathbf{G}_2}(\mathbf{k}) =\notag \\ &\sum_{\mathbf{p}\mathbf{G}n}\frac{V(\mathbf{p}-\mathbf{k}+\mathbf{G}_1-\mathbf{G}_2)}{N_k} u^*_{\mathbf{p}\mathbf{G}-\mathbf{G}_1+\mathbf{G}_2 \beta n}u_{\mathbf{p}\mathbf{G}\alpha n} f_0(\epsilon_{\mathbf{p}n})
\end{align}
where $N_k$ is the number of points within the cutoff, $f_0(\epsilon) = (e^{\beta(\epsilon-\mu)} + 1)^{-1}$ is the Fermi-Dirac distribution, with $\beta = 1/k_BT$ being the inverse temperature and $\mu$ the chemical potential.

The order parameters given above, which among others represent various forms of excitonic order, enter the mean field Hamiltonian through:
\begin{equation}
H_{\mathrm{HF}} = \sum_{\mathbf{k},\alpha,\beta} \sum_{\mathbf{G}_1, \mathbf{G}_2} (\Delta_{\mathrm{H}\alpha\beta}^{\mathbf{G}_1\mathbf{G}_2} - \Delta_{\mathrm{F}\alpha\beta}^{\mathbf{G}_1\mathbf{G}_2}(\mathbf{k})) c^\dagger_{\mathbf{k}\mathbf{G}_1\alpha} c_{\mathbf{k}\mathbf{G}_2\beta}
\end{equation}

The new $H_\mathrm{MF}$ is then again diagonalized and the whole procedure is repeated until convergence is achieved, which is monitored by the change in the average value of the order parameters $\Delta_{\mathrm{H}\alpha\beta}^{\mathbf{G}_1\mathbf{G}_2}$ and $\Delta_{\mathrm{F}\alpha\beta}^{\mathbf{G}_1\mathbf{G}_2}(\mathbf{k})$, with the calculation ending when the difference between each step is smaller than $10^{-14}$.

The appearance of various excitonic phases is established by calculation of several different quantities. The overall presence of excitonic condensate with $\mathbf{q}_c$ momentum is determined through:
\begin{equation}
\Delta_\mathrm{exc} = \sqrt{\frac{1}{N_k} \sum_{\mathbf{k}\mathbf{G}\alpha\beta n} |u^*_{\mathbf{k}\mathbf{G}\alpha n} u_{\mathbf{k}\mathbf{G}+\mathbf{q}_c \beta n} f_0(\epsilon_{\mathbf{k} n})|^2}
\end{equation}
The time-reversal breaking spin spiral and spin density wave components are characterized using the Fourier components of spin density that correspond to $\mathbf{q}_c$ ordering vector:
\begin{equation}
\rho^s_i = \frac{1}{N_k} \sum_{\mathbf{k}\sigma\sigma'an} s_{i, \sigma\sigma'} u^*_{\mathbf{k}\mathbf{0}\sigma a n} u_{\mathbf{k}\mathbf{q}_c \sigma' a n} f_0(\epsilon_{\mathbf{k} n})
\end{equation}
Expressing the spin density as a vector $\bm{\rho}^s$ we can then give the expressions that separate the spin spiral and spin density wave components:
\begin{equation}
\rho^\mathrm{SDW} = \sqrt{2 |\bm{\rho}^s\cdot \bm{\rho}^s|}, \quad \rho^\mathrm{SS} = \sqrt{2 |\bm{\rho}^s|^2} - \rho^\mathrm{SDW}
\end{equation}

The exact phase obtained as the Hartree-Fock ground state depends on several factors, most important of which is the screened interaction strength $V_0$. This parameter is determined by the device configuration, in particular the spacing between the metallic gates that are used to control charge density within the device and the gate insulator material itself. Through the self-consistent calculation we obtain results in agreement with the previous calculations with \cite{jiaEvidenceMonolayerExcitonic2022} and without \cite{Kwan2021} time-reversal symmetry, reproducing the phase diagram that contains spin spiral and spin density wave in the latter case. Representative examples of mean-field band structures are presented in Fig.~\ref{fig:bands}. In panel (a) we show that when time-reversal symmetry is enforced, the exciton condensate forms and the gap opens up, but the bands remain doubly degenerate. The state is topological in nature as will be explicitly demonstrated by the presence of the edge states. In panel (c), the band structure for time-reversal breaking state is shown. Again, the exciton condensate formation leads to gap opening, but now the band degeneracy in the proximity of the gap is lifted due to the formation of the spin spiral state. As this degeneracy-lifting is small, the system will still retain some of the quantum spin Hall effect features, but will no longer be robust to perturbations. 

\begin{figure*}
    \centering
    \includegraphics[width=0.8\linewidth]{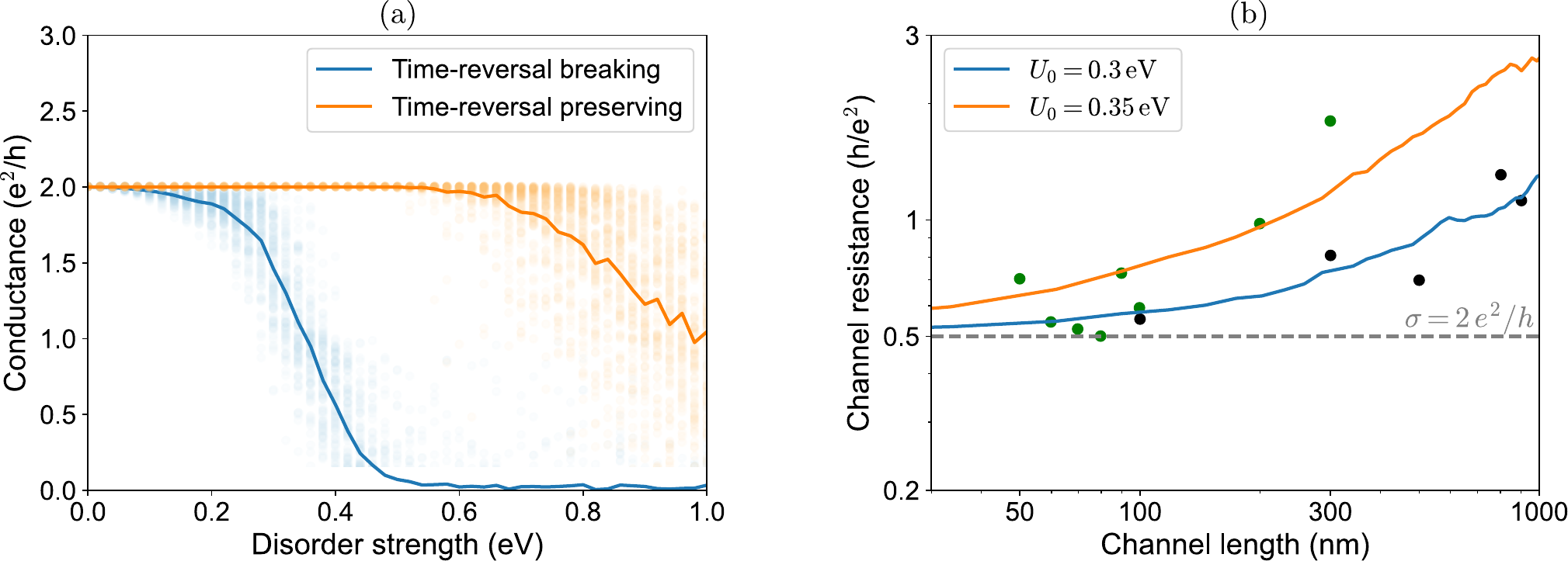}
    \caption{Quantum transport simulation results. (a) Disorder strength $U_0$ dependence of conductance for the case of time-reversal breaking and preseving exciton condensates. The solid lines show average over 100 disorder realizations, with each realization shown as a separate set of points. (b) The comparison between the experimental data as reported in \cite{wuObservationQuantumSpin2018a} and quantum transport simulation averaged over disorder realizations. The experimental data from two different devices and channel lengths 50-1000 nm demonstrates increasing channel resistance for longer edge channels, suggesting lack of topological protection of the quantum spin Hall edge states. This observation is consistent with the simulation result, which can reproduce the observed length dependence and sample variation with relatively small changes to disorder strength value $U_0$.}
    \label{fig:quantum_transport}
\end{figure*}

For the purpose of quantum transport simulation with disorder we need to convert these Hamiltonian terms to their real space counterparts. While the unit cell of the original tight-binding Hamiltonian consisted of a single site with four spin-orbitals, when excitonic order with finite $q$ arises, the unit cell has to increase correspondingly to allow for the spatial modulation of the charge and spin densities.   Since we have chosen the lattice constant so that the postulated excitonic order wave vector is commensurate with the original lattice, we can simply extend the original model by including $L_x$ sites in the extended unit cell. The chosen value of $L_x=7$ enables a reasonable approximation for the modulation of densities within the unit cell while keeping the computational complexity in check. At the same time, we can easily convert the Hartree-Fock order parameters to real space hoppings. As a result of a Fourier transform with $c^\dagger_{\mathbf{k}\alpha} = 1/\sqrt{N} \sum_{\mathbf{r}_i}e^{i \mathbf{k}\cdot\mathbf{r}_i}c^\dagger_{\mathbf{r}_i}$ we obtain hoppings between $\mathbf{r}_i$ and $\mathbf{r}_j$ sites of the lattice:

\begin{equation}
\Delta_{\alpha\beta}(\mathbf{r}_i, \mathbf{r}_j) = \sum_{\mathbf{k}\mathbf{G}_1\mathbf{G}_2} \Delta_{\alpha\beta}^{\mathbf{G}_1\mathbf{G}_2}(\mathbf{k}) e^{i \mathbf{k}\cdot\mathbf{r}} e^{i((\mathbf{G}_1-\mathbf{G}_2)\cdot \mathbf{r}_j+\mathbf{G}_1\cdot\mathbf{r})} 
\end{equation}
where $\Delta_{\alpha\beta}^{\mathbf{G}_1\mathbf{G}_2}(\mathbf{k}) = \Delta_{\mathrm{H}\alpha\beta}^{\mathbf{G}_1\mathbf{G}_2} - \Delta_{\mathrm{F}\alpha\beta}^{\mathbf{G}_1\mathbf{G}_2}(\mathbf{k})$ and $\mathbf{r} = \mathbf{r}_i - \mathbf{r}_j$. Similarly to the discretization of the continuum model, we use hoppings up to ($\pm$3, $\pm$ 3) in $x$ and $y$ directions. While the Fourier decomposition of $\Delta_{\alpha\beta}^{\mathbf{G}_1\mathbf{G}_2}(\mathbf{k})$ includes an infinite number of $\Delta_{\alpha\beta}(\mathbf{r}_i, \mathbf{r}_j)$, in practice for the system under study such a limited number of terms reproduces the Hartree-Fock potential with sufficient accuracy, partially due to the choice of lattice constant that results in an order that is commensurate with the initial lattice.

With a real-space model defined as above, we can now set up the quantum transport simulation to study the effect of the time-reversal-breaking excitonic condensate on the helical edge states of the quantum spin Hall state. To perform all the quantum transport calculations we use the Kwant package \cite{grothKwantSoftwarePackage2014}. We restrict the model to a ribbon with a finite width in $y$ direction, while still retaining the translational invariance in the $x$ direction, along the spin density modulation coming from the spin spiral state. Due to this translational invariance we can still label the quantum states by their momentum in $x$ direction and calculate the spectrum within the 1D Brillouin zone (which is indicated by primes in the point labeling) of the ribbons. Such spectra are presented in Fig.~\ref{fig:bands}, both for time-reversal preserving and breaking cases. In panel (b), the time-reversal preserving case, the helical edge states resulting from the topological nature of the excitonic insulator are clearly visible. However, in panel (d) even though the time-reversal symmetry is broken, the remainder of helical edge states is still visible in the gap. These states are partially gapped, but because the magnitude of symmetry breaking terms is not large, there are energy windows within the gap for which a pair of edge states is present and in the absence of any disorder they would contribute 2$e^2/h$ to the longitudinal conductance of the ribbon.

In a standard Landauer-Buttiker calculation fashion, we attach the semi-infinite leads to the opposite ends of the ribbon that extend in the $x$ direction. We then introduce additional random on-site potential to the tight-binding model within the central scattering region to model the disorder that preserves time-reversal symmetry:

\begin{equation}
H_\mathrm{dis} = \sum_{\mathbf{r}_i, \alpha} U(\mathbf{r}_i) c^\dagger_{\mathbf{r}_i, \alpha} c_{\mathbf{r}_j, \alpha}
\end{equation}

The random on-site values $U(\mathbf{r}_i)$ are taken from a uniform distribution over the range $[-U_0/2, U_0/2]$, where we call $U_0$ the disorder strength. We calculate the scattering matrix of the system for $n_\mathrm{avg}=100$ independent disorder realizations and then average the conductances over these realizations. In the usual circumstances, the quantum spin Hall edge states are robust with regards to such a disorder and the breakdown of the quantized sample conductance happens only for extremely large disorder strengths. This remains true even in the presence of the excitonic condensate which preserves time-reversal symmetry, as demonstrated in Fig.~\ref{fig:quantum_transport}(a). The conductance remains precisely quantized for much of the investigated disorder strength range and the leading source of the quantization breakdown is the coupling between the opposite edges of the sample. This is evident as the conductance curve is sensitive to the width of the ribbon, with the disorder strength needed to decrease the conductance increasing for wider ribbons. At the same time, when disorder is chosen within the quantized plateau, the result doesn't depend on the length of the ribbon, revealing robustness characteristic of the topological transport. Moreover, the variance displayed among the different disorder realizations shows that conductance in that case is very unstable, being dependent on whether the disordered potential forms a path connecting to the opposite edge.

However, once the time reversal symmetry breaking excitonic condensate is formed, even a small scalar disorder causes a deviation from the perfect value, as exemplified in Fig.~\ref{fig:quantum_transport}(a). In contrast to the time-reversal preserving case, the decrease from 2$e^2/h$ value is immediate. Moreover, the disorder strength dependence is insensitive to the width scaling, indicating that the backscattering processes occur within the same edge. The variance among disorder realizations is also much smaller then previously, confirming that the backscattering is not depedent on accidental appearance of a pathway across the device. At the same time, increasing the length of the ribbon decreases the conductance for all disorder strengths, revealing the lack of the topological protection. 

To further demonstrate this, and to relate our simulations to the experimental data, we calculate the channel resistance of the edge states with increasing ribbon length, keeping all the other Hamiltonian parameters, including the disorder strength $U_0$, constant. The results are presented in Fig.~\ref{fig:quantum_transport}(b), where a comparison between the simulation and experimentally obtained resistance values \cite{wuObservationQuantumSpin2018a} is made. The experimentally studied edge lengths range from 50 to almost 1000 nm, which we also use as the ribbon lengths in our calculations. We include calculations with two different, but comparable disorder strengths. The smaller value, $U_0 = 0.3$\,eV reproduces the device data represented by the black points, while the larger disorder strength $U_0 = 0.35$\,eV is well-suited for the device represented by the green points. This demonstrates that the variability between the devices can be explained by relatively small change in disorder strength that can realistically be expected in experimental conditions. In both cases, the simulations mirror the experimental data for almost two orders of magnitude in channel length, suggesting that the breakdown of quantization of edge conductance as observed in WTe$_2$ could be explained by a bulk time-reversal-breaking excitonic condensate.

\section{Transport from the edge theory}\label{SecEdgeTransport}
To better understand the mechanism behind the breakdown of quantization in the presence of the time-reversal-breaking excitonic condensate, we use an analytical model based on the Luttinger liquid picture. Such a model can be applied to study the helical edge states that arise in the quantum spin Hall (QSH) insulator under open boundary conditions. The helical edge state consists of two degenerate counterpropagating electron states with opposite spins, related by the time reversal symmetry, each contributing a quantized zero temperature conductance $G_0 \equiv e^2/h$ in opposite directions. Even though the quantized conductance $G_0$ is protected by time reversal symmetry, there exists a multitude of proposals that can explain the breakdown of the quantization as discussed in the Introduction. In terms of time-reversal breaking mechanisms, previous works suggested backscattering from magnetic impurities as one of the most prominent. However, in many systems such as WTe$_2$ the source of such impurities is not obvious and thus we look towards a different, intrinsic source of time-reversal breaking, namely the excitonic condensate. In this section, following Ref.~\cite{Hsu2017,Hsu2018,Hsu_2021}, we compute the two terminal conductance for a QSH insulator slab with bulk spin spiral order. While in the referenced works the justification for the appearance of such an order is the spontaneous arrangement of nuclear spins, here the spin spiral is purely electronic as shown in the Hartree-Fock calculations, giving hope for a more pronounced effect at higher temperatures. Although the spin spiral order breaks the time reversal symmetry locally, when averaged over the whole period of the spiral the symmetry is still preserved, resulting in a lack of gap opening at the Dirac point of the edge state dispersion. Therefore, we need to also take into account scattering on random non-magnetic impurites to observe the effect on conductance. To put this on more concrete terms, we use the bosonization technique and compute the deviation from the perfect conductance quantum $G_0$ below.

\begin{figure}
    \centering
    \includegraphics[width=1\linewidth]{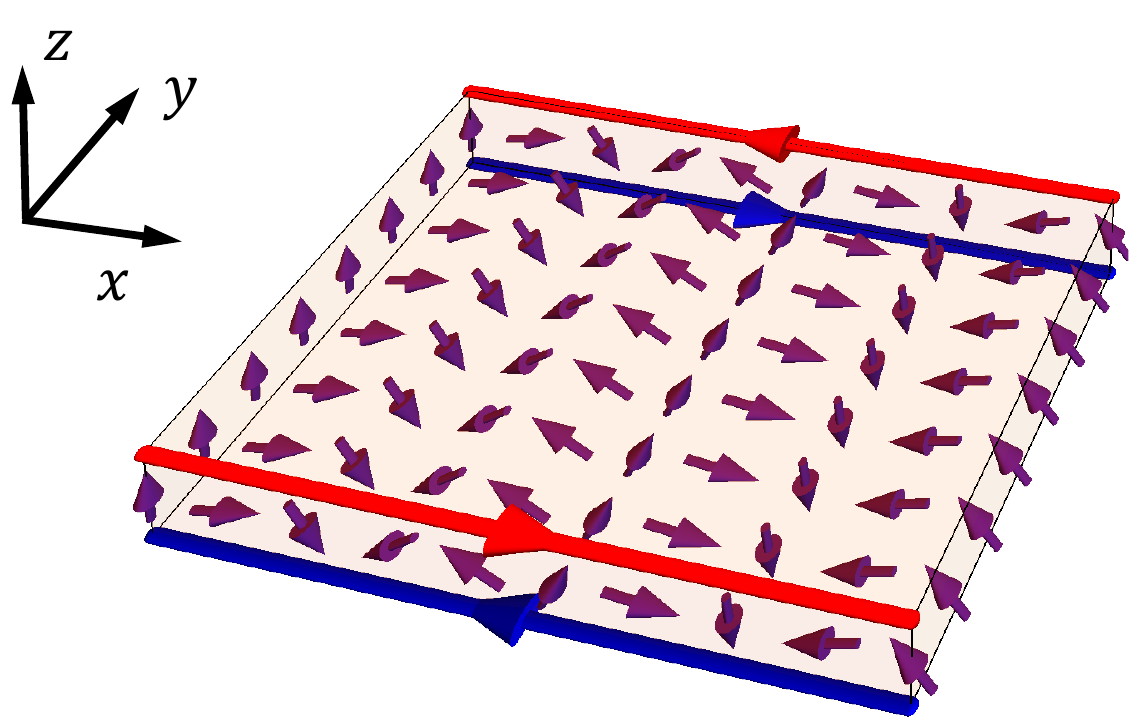}
    \caption{Illustration of spin spiral order in the quantum spin Hall phase. The single layer of WTe$_{2}$ (blue cuboid) is in a quantum spin Hall effect phase before the appearance of the bulk spin spiral order (purple arrows). The red and blue arrows on the boundary stand for the helical edge states from the original quantum spin Hall effect. }
    \label{Illustration}
\end{figure}

\subsection{Fermionic model}
\begin{figure}
    \centering
    \includegraphics[width=0.5\linewidth]{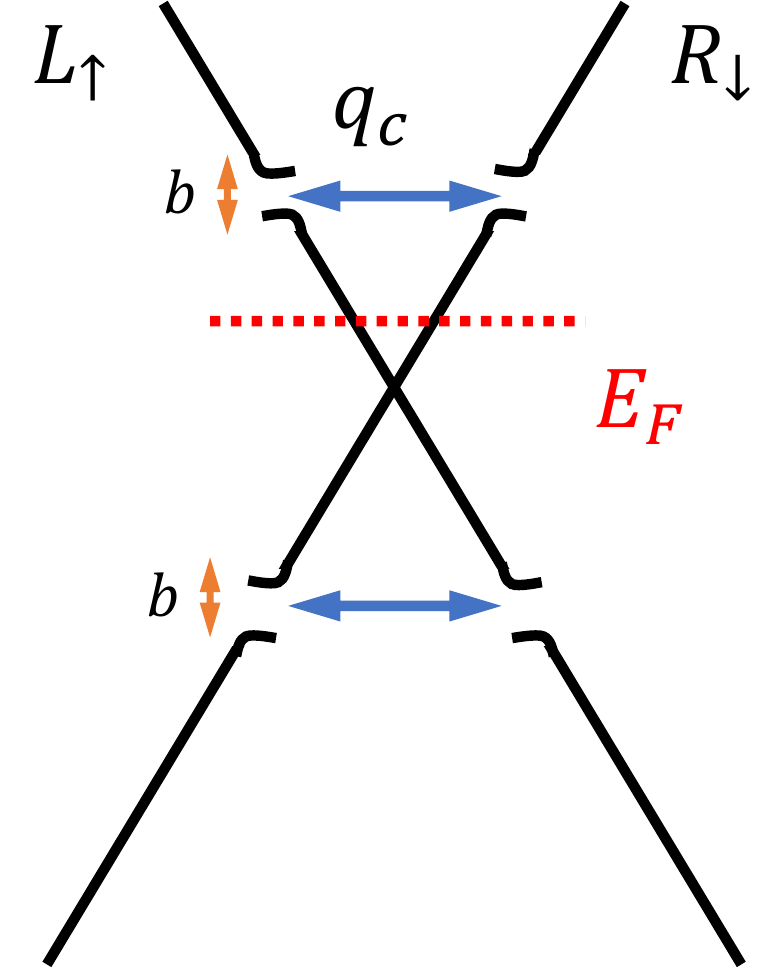}
    \caption{The spin spiral will not gap out the Dirac cone generically in the absence of disorder. Instead, it will open a gap on the edge of the Brillouin zone at $\pm q_c/2$, where the effective real-space periodicity is defined by the spiral.}
    \label{Edge_Gap}
\end{figure}

We consider the setup for a single layer of QSH insulator WTe${}_2$ as shown in Fig.~[\ref{Illustration}], in the form of a ribbon with open boundary conditions along $y$ direction. This results in helical edge states (denoted by the red and blue arrows) on the upper and lower boundaries. Based on the results of Hartree-Fock calculations, we further assume the ground state of the QSH insulator has bulk spin-spiral order, denoted by the violet arrows. Such a bulk spin spiral preserves the time reversal symmetry only when spatially averaged over the period of the spiral along the $x$ direction. As the edge state decay length is very short in WTe$_2$ \cite{bieniek2022theory}, we can neglect any coupling between the states located at the opposite boundaries. Focusing thus on states along a single boundary, we can write down the Hamiltonian for helical edge states as:
\begin{equation}\label{Helical_Edge_Hamiltonian}
	H_{\rm hel} = H_{\rm kin} + H_{\rm ee} + H_{\rm m}+H_{\rm imp},
\end{equation}
where $H_{\rm kin}$ and $H_{\rm ee}$ are the kinetic energy and electron-electron interactions, respectively. The term $H_{\rm m}$ is the effective magnetic field from the bulk excitonic spin spiral, and $H_{\rm imp}$ denotes the non-magnetic impurities. We further assume that the helical edges are formed by a right-moving mode with spin down ($R_\downarrow$) and a left-moving mode with spin up ($L_\uparrow$). The kinetic energy for the helical edge states reads:
\begin{equation}
	H_{\rm kin} = - i\hbar v_F \int [dx] [R^\dagger_\downarrow(x) \partial_x R_\downarrow(x) - L^\dagger_\uparrow (x) \partial_x L_\uparrow(x)],
\end{equation}
with the Fermi velocity $v_F$. When time reversal symmetry is preserved, the electron-electron interaction $H_{\rm ee} = H^2_{\rm ee} + H^4_{\rm ee}$ contains only forward scattering $H_{\rm ee}^2$ and chiral interaction $H^4_{\rm ee}$:
\begin{subequations}
\begin{align}
H_{\rm ee}^2 &\!=\! g_2 \int[dx] R_\downarrow^\dagger(x) R_{\downarrow}(x) L_\uparrow^\dagger(x) L_\uparrow(x), \\
H_{\rm ee}^4 &\!=\! \frac{g_4}{2} \int[dx] [(R^\dagger_\downarrow(x) R_\downarrow(x))^2 + (L_\uparrow^\dagger(x)L_\uparrow(x))^2].
\end{align}
\end{subequations}
The $H_{\rm m}$ captures coupling between the spin density of the edge states and the effective magnetic field induced by the bulk spin spiral order:
\begin{equation}
	H_{\rm m} = \int [dx]\sum_{ss^\prime} \psi^\dagger_s(x) ({\bm b}(x) \cdot{} {\bm \sigma}_{ss^\prime}) \psi_{s^\prime}(x),
\end{equation}
where  ${\bm \sigma}$ is a vector of Pauli matrices and $\psi(x) = (L_\uparrow,R_\downarrow)^{\rm T}$. The spatial dependence of the coupling to effective magnetic field ${\bm b}(x)$ can be deduced from the Hartree-Fock mean field and is given by:
\begin{equation}
	{\bm b}(x) = b[\hat e_x \cos(q_c x) + \hat e_z \sin(q_cx)].
\end{equation} 
Here $2\pi/q_c$ denotes the period of the bulk spin spiral. In the non-interacting clean limit, i.e., $g_2 = g_4 = 0$ and $H_{\rm imp} = 0$, we can transform Eq.~[\ref{Helical_Edge_Hamiltonian}] into Fourier space:
\begin{equation}\label{Helical_Non_Interacting}
\begin{split}
    H_{\rm hel}^\prime = &\sum_{k,s,s^\prime} \bigg{[} \psi^\dagger_s(k) (\hbar v_F k \sigma^z_{ss^\prime}) \psi_{s^\prime}(k)\, +\\ & \frac{b}{2}\psi_{s}^\dagger(k+q_c)(\sigma_{ss^\prime}^x - i \sigma^z_{ss^\prime})\psi_{s^\prime}(k) +H.c.   \bigg{]}.
\end{split}
\end{equation}
Note that Eq.~[\ref{Helical_Non_Interacting}] couples the right movers and left movers separated by a momentum difference $\pm q_c$. Thus the energy spectrum of the Hamiltonian of Eq.~[\ref{Helical_Non_Interacting}] is gapless at $k = 0$, but gaps open for states at momenta $\pm q_c/2$, as shown in Fig.~[\ref{Edge_Gap}]. The lack of gap opening at $k=0$ reflects that $H_{\rm m}$ preserves the time-reversal symmetry when averaged over real space.

Since we are only interested in the impact of the spin spiral on transport properties, we can neglect the forward scattering components of $H_\mathrm{m}$, which are the terms proportional to $\sigma_z$. Therefore, we can now focus only on the backscattering part, which written in terms of $L_\uparrow(x)$ and $R_\downarrow(x)$ reads:
\begin{equation}
    H_{\rm m}^b \!=\! \int [dx]\frac{b}{2}\sum_{\delta = \pm} [L_\uparrow^\dagger(x) R_\downarrow(x) e^{iq_c^\delta x} + R_\downarrow^\dagger(x) L_\uparrow(x) e^{-iq_c^\delta x} ],
\end{equation}
with $q_c^\delta = \pm q_c$. We further model the impurity Hamiltonian as:
\begin{equation}
    H_{\rm imp} = \int [dx] V_{\rm imp}(x) [R^\dagger_\downarrow(x) R_\downarrow(x) + L_\uparrow^\dagger(x)L_\uparrow(x)],
\end{equation}
where the Gaussian random potential $V_{\rm imp}(x)$ satisfies $\overline{V_{\rm imp}(x)V_{\rm imp}(x^\prime)} = M_{\rm imp}\delta(x-x^\prime)$, with $\overline{\cdots}$ denoting the averaging over the random potential. The impurity strength $M_{\rm imp} = \hbar^2 v_F^2 / (2\pi \lambda_{\rm mfp})$ here is defined by the mean free path $\lambda_{\rm mfp}$ of the 2D QSH insulator bulk~\cite{Hsu2017,Hsu2018,Hsu_2021}.

\subsection{Schrieffer-Wolff transformation}
\begin{figure}
    \centering
    \includegraphics[width=0.8\linewidth]{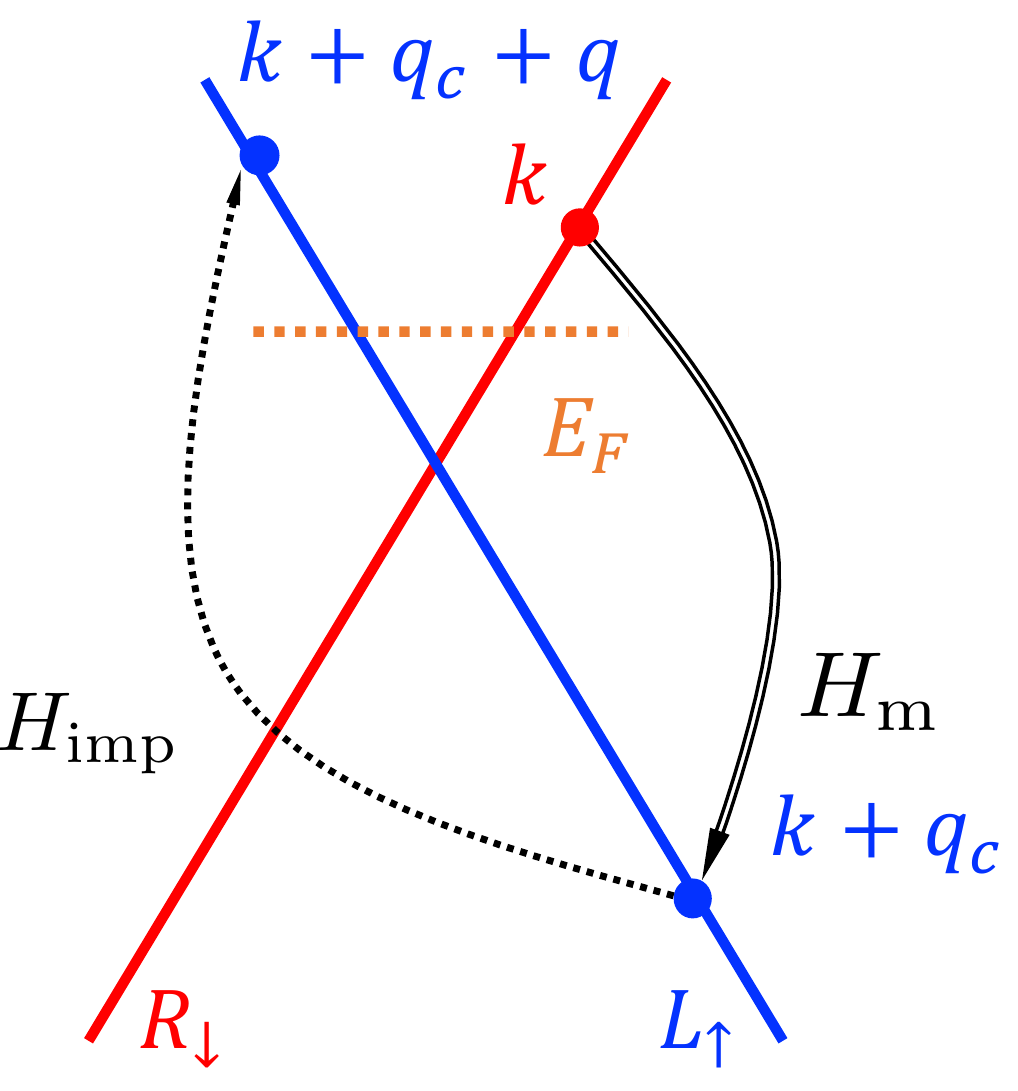}
    \caption{Spin spiral assisted backscattering on non-magnetic impurities. The initial right moving state with momentum $k$ is first scattered by the spin spiral to a left mover with momentum $k+q_c$ and then forward scattered by an impurity to a left moving state with momentum $k+q_c+q$. At low temperatures, the leading scattering processes will have the initial and final state in the vicinity of the Fermi energy $E_F$}
    \label{Spiral_Assited_Scattering}
\end{figure}

We can now derive the combined effect of the random impurities and the spin spiral arising from the excitonic condensate by performing a Schrieffer-Wolff transformation. The full Hamiltonian with the two perturbations included is $H = H_{\rm el} + \delta V$, where  $\delta V = H_{\rm imp} + H_{\rm m}$. This can be transformed into momentum space using:
\begin{equation}
	R_\downarrow(x) = \frac{1}{\sqrt{L}}\sum_k e^{ikx}R_\downarrow(k), \quad L_\uparrow(x) = \frac{1}{\sqrt{L}}\sum_k e^{ikx}L_\uparrow(k), 
\end{equation}
such that we have kinetic energy diagonal in momentum space
\begin{equation}
\begin{aligned}
	H_{\rm kin} = \sum_k \hbar v_F k [R_\downarrow^\dagger(k) R_\downarrow(k) -L_\uparrow^\dagger(k)  L_\uparrow (k) ]. 
\end{aligned}
\end{equation}
The impurity contributes arbitrary momentum shift in the forward scattering process
\begin{equation}
			H_{\rm imp} = \sum_{k,q}\frac{V_{\rm imp}(q)}{L}  \left( R_\downarrow^\dagger(k + q) R_\downarrow(k) + L^\dagger_\uparrow(k + q)L_\uparrow(k) \right)
\end{equation}
and the coupling to the spin spiral reads:
\begin{equation}
	\begin{aligned}
			H_{\rm m} = \frac{b}{2L} \sum_{k,\delta}[L^\dagger_\uparrow(k+q^\delta_c)R_\downarrow(k) + R_\downarrow^\dagger(k+q^\delta_c)L_\uparrow(k)]
	\end{aligned}
\end{equation}
which changes a right to a left mover and vice versa together with a momentum shift $\pm q_c$ determined by the period of the spin spiral. The forward scattering from $H_{\rm imp}$ on its own does not change the conductivity and at low temperatures, the initial and final states in backscattering process should come from the vicinity of the Fermi level. Therefore, neither $H_{\rm imp}$ nor $H_{\rm m}$ in 1D can affect conductivity on their own. However, a combined spin spiral and impurity scattering can lead to the relaxation of the current through a second-order process, an example of which is illustrated in Fig.~[\ref{Spiral_Assited_Scattering}]. This second-order effect, where the $H_{\rm m}$ first backscatters a state with momentum $k$ to $k+q_c$ ($k-q_c$), and then $H_{\rm imp}$ brings the state back to the region near the Fermi level through the forward scattering with the disorder potential. We see that the scattering with combination of the spin spiral and non-magnetic disorder can be viewed as the scattering with an effective magnetic disorder, though the former pattern still respects time reversal symmetry on average. We term this kind of the current relaxation mechanism as the spin-spiral assisted backscattering.

The intuitive picture of spin-spiral assisted backscattering presented above can be captured by the Schrieffer-Wolff transformation. For simplicity, we first consider the non-interacting case where $H_{\rm el} = H_{\rm kin}$.  
\begin{equation}
\begin{aligned}
	H^\prime &= e^S H e^{-S} =H_\mathrm{kin} + \delta V + [S,H_\mathrm{kin}] + [S,\delta V] \\
 &+ \frac{1}{2} [S,[S,H_\mathrm{kin}]] + \frac{1}{2}[S,[S,\delta V]] + \cdots.
\end{aligned}
\end{equation}
We can now choose $S$ such that the transformed Hamiltonian $H^\prime$ does not depend on perturbation couplings to linear order by fulfilling the condition $\delta V + [S,H_\mathrm{kin}] = 0$. Since $H_{\rm kin}$ is diagonal in momentum space, we can write down the matrix element $S_{\alpha \beta}$ between the two eigenstates $\alpha$ and $\beta$ of $H_{\rm kin}$:
\begin{equation}
    S_{\alpha \beta} = \frac{\delta V_{\alpha \beta}}{E_\alpha-E_\beta}.
\end{equation}

Substituting this back into $H^\prime$, we obtain:
\begin{equation}\label{SW_In_Details}
   H^\prime_{\alpha \beta} \!=\! E_\alpha \delta_{\alpha \beta} + \frac{1}{2}\sum_\gamma\bigg{(} \frac{ \delta V_{\alpha \gamma} \delta V_{\gamma \beta} }{E_\alpha - E_\gamma} + \frac{ \delta V_{\alpha \gamma} \delta V_{\gamma \beta} }{E_\gamma - E_\beta} \bigg{)} + O (\delta V^3).
\end{equation}
We can now consider how the original Hamiltonian acts within the subspace of the three states presented in Fig.~\ref{Spiral_Assited_Scattering}. In the basis of $\Psi = (R_\downarrow(k),L_\uparrow(k+q_c),L_\uparrow(k+q_c+q))/L$, we have:
\begin{equation}
    \begin{aligned}
        h(k,q) \!=\! \Psi^\dagger \begin{pmatrix}
        \hbar v_F k & 0 & b/2 \\
        0 & -\hbar v_F(k+q_c + q) & V_{\rm imp} \\
        b/2 & V_{\rm imp} & -\hbar v_F (k+q_c) 
        \end{pmatrix} \Psi.
    \end{aligned}
\end{equation}
Using these matrix elements in combination with Eq.~[\ref{SW_In_Details}], we find:
\begin{equation}
    \Delta h(k,q) \!=\!   \frac{b}{4}\bigg{(} \frac{V_{\rm imp}(q)}{\hbar v_F(2k + q_c) } - \frac{V_{\rm imp}(q)}{\hbar v_F q } \bigg{)}.
\end{equation}
Now we need to take into account the conservation of energy in the whole scattering process. The majority of the contribution in the low temperature limit is given by the states with an identical energy for the initial and final states around the Fermi level: with initial $k\sim k_F$, these energies are $\hbar v_F k_F$ and $- \hbar v_F (k_F+q_c - q)$, leading to $q = -2k_F -q_c + k^\prime$. In such limit, we have:
\begin{equation}
    \Delta h (k^\prime) \approx \frac{b}{2} \frac{V_{\rm imp}(-2k_F - q_c + k^\prime)}{\hbar v_F (2k_F + q_c)}.
\end{equation}
Similar procedure follows for the other possible scattering processes and in the end the effective back scattering reads:
\begin{equation}\label{Effective_SS_Impurity}
    \begin{aligned}
        H_{\rm eff} &= \sum_{k,k^\prime,\delta}[  V^\delta(k^\prime) L^\dagger_\uparrow(k- 2k_F+k^\prime) R_\downarrow(k) + {\rm H.c.}],
    \end{aligned}
\end{equation}
where $V^\pm(k^\prime) = b V_{\rm imp}(-2k_F \mp q_c + k^\prime)/(2L^2\hbar v_F (2k_F \pm q_c))$.

The above results hold even in the interacting case, provided that $g_2 = g_4$~\cite{Hsu2018}. While Eq.~[\ref{Effective_SS_Impurity}] looks very much similar to Eq.~[D5] in Ref.~\cite{Hsu2018}, the physical interpretation is quite different. In Ref.~\cite{Hsu2018}, the spin spiral arises from the RKKY coupling of nuclear spins, whereas in our case it comes from the time-reversal breaking excitonic condensate. The nuclear spin spiral period is thus directly related to $k_F$ and changes with chemical potential, while the excitonic spiral is determined by the momentum space separation between the electron and hole pockets in the band structure and is fixed.

The Eq.~[\ref{Effective_SS_Impurity}] can be transformed into real space by the inverse Fourier transformation after which the Hamiltonian reads:
\begin{equation}
    \begin{aligned} 
        H_{\rm eff} &= \int[dx] [ \xi(x)L_\uparrow^\dagger(x) R_\downarrow(x)  + \xi^*(x) R_\downarrow^\dagger(x) L_\uparrow(x)],
    \end{aligned}
\end{equation}
with:
\begin{equation}
    \xi(x) =b \frac{2k_F \cos (q_c x) +i q_c \sin (q_c x)}{\hbar v_F(4k_F^2 - q_c^2)}V_{\rm imp}(x),
\end{equation}
The correlation function of $\xi(x)$ averaged over the disorder realizations can be deduced from the correlation of $V_\mathrm{imp}(x)$ to be:
\begin{equation}
    \overline{\xi(x) \xi^*(x^\prime)} = M_{\rm ss} \delta(x-x^\prime),
\end{equation}
with
\begin{equation}
   M_{\rm ss} =  M_{\rm imp} \frac{b^2(4k_F^2 + q_c^2)}{2\hbar^2 v_F^2(4k_F^2 -q_c^2)^2},
\end{equation}
where we used the fact that $\overline{\cos^2(q_cx)} =\overline{\sin^2(q_cx)} =1/2$.

\subsection{Bosonization result for transport propertied}
With the effective spin spiral assisted backscattering Hamiltonian derived, we can determine its impact on transport properties using bosonization, which is especially well suited for studying two-terminal transport in disordered 1D quantum system~\cite{Safi1995,Maslov1995A,Maslov1995B,Hsu_2021,Hsu2017,Hsu2018,Kwan2021,Vidal1999,Vidal2001,Kainaris2014}. The chiral component in Eq.~[\ref{Helical_Edge_Hamiltonian}] can be expressed in terms of the bosonic field $(\theta,\phi)$~\cite{Hsu_2021,Hsu2017,Hsu2018,Kwan2021,Vidal1999,Vidal2001,Kainaris2014}:
\begin{equation}
\begin{split}
    R_\downarrow(x) &= \frac{U_R}{\sqrt{2\pi \alpha}} e^{ik_Fx} e^{i[-\phi(x) + \theta(x)]}, \\ L_\uparrow(x) &= \frac{U_L}{\sqrt{2\pi \alpha}} e^{-ik_Fx} e^{i[\phi(x) + \theta(x)]},
\end{split}
\end{equation}
where $U_{R/L}$ is the Klein factor and $\alpha = \hbar v_F /\Delta_b$ is the short-distance cutoff, which is associated with the high-energy cutoff set by the bulk gap $\Delta_b$.

 With the above definitions, the helical Hamiltonian without $H_{\rm m}$ can be bosonized in a standard way as:
\begin{equation}
	H_{\rm kin} + H_{\rm ee} = \frac{\hbar u}{2\pi} \int [dx] \bigg{[} \frac{1}{K} (\partial_x \phi)^2 + K(\partial_x \theta)^2 \bigg{]},
\end{equation}
where the velocity $u$ and the interaction parameter $K$ are given by:
\begin{equation}
\begin{aligned}
	u&= \bigg{[} \bigg{(} v_F + \frac{g_4}{h} \bigg{)}^2 - \bigg{(} \frac{g_2}{h}\bigg{)}^2 \bigg{]}^{1/2}, \\
 K &\equiv \bigg{(} \frac{h v_F + g_4 - g_2}{h v_F + g_4 + g_2}\bigg{)}^{1/2}.
 \end{aligned}
\end{equation}
To find the bosonized form of the effective backscattering Hamiltonian $H_{\rm eff}$ averaged over disorder, we utilize the replica method, and similarly to Ref.~\cite{Hsu2018}, we obtain the effective backscattering action:
\begin{equation}\label{Deviation}
\begin{aligned}
    \frac{\delta S_{\rm ss}}{\hbar} \!=\! &- \frac{M_{\rm ss}}{(2\pi \hbar a)^2} \int_{u|\tau - \tau^\prime| > a} dx d\tau d \tau^\prime \\
    &\times \cos[2 \phi(x, \tau) - 2 \phi(x, \tau^\prime)].
\end{aligned}
\end{equation}
In the absence of disorder, so long as the Fermi level is not placed within the gap (see in Fig.~[\ref{Edge_Gap}]), the Hamiltonian Eq.~[\ref{Helical_Edge_Hamiltonian}] is similar to the edge of time reversal invariant QSH insulator, which has a quantized zero-temperature conductance  $G = G_0 \equiv e^2/h$ (equivalently, resistance $R_0 = h/e^2$) per edge. Following the results of Ref.~\cite{Hsu_2021}, we find the increase of the resistivity $R_{\rm ss}$ due to the combination of bulk spin spiral and non-magnetic disorder depends on the relative magnitude of three possible physical cutoffs. These cutoffs are the edge length $L$, the thermal length $\lambda_T \equiv \hbar u / (k_BT)$, and the localization length $\xi_{\rm ss}$. We define the dimensionless coupling constant, 
\begin{equation}
    D_{\rm ss} \equiv \frac{2a M_{\rm ss}}{(\pi \hbar^2 u^2)} ,
\end{equation}
and $u = v_F/K$. If the edge length is the shortest among all these scales, $L < \lambda_T, , \xi_{\rm ss}$, the correction to edge resistance is:
\begin{equation}
    \delta R_{\rm ss}(L) \propto R_0 \frac{M_{\rm ss}L}{\hbar^2 v_F^2} L^{2-2K} =R_0 \frac{\pi D_{\rm ss}}{2K^2} \bigg{(} \frac{L}{a} \bigg{)}^{3-2K}.
\end{equation}
Secondly, in the limit of high temperatures where $\lambda_T < L, , \xi_{\rm ss}$, we get:
\begin{equation}
   \delta R_{\rm ss}(T) \propto R_0 \frac{M_{\rm ss}L}{\hbar^2 v_F^2} \lambda_T^{2-2K} = R_0 \frac{\pi D_{\rm ss}L}{2K^2a} \bigg{(} \frac{Kk_BT}{\Delta} \bigg{)}^{2K-2}.
\end{equation}
Finally, if $\xi_{\rm ss}< L, \lambda_T$, the RG flow reaches the strong coupling regime, so the edge states are gapped, displaying a thermally activated resistance:
\begin{equation}
    \delta R_{\rm ss}(T)\propto R_0 \frac{\pi D_{\rm ss} L}{2K^2 a} e^{\Delta_{\rm ss}/(k_BT)},
\end{equation}
with the gap $\Delta_{\rm ss} = \Delta (2K D_{\rm ss})^{1/(3-2K)}$.

\section{Conclusion}\label{SecConclusion}
In conclusion, we provided a mechanism for the breakdown of the perfect quantized conductance quantization in WTe$_2$ due to the formation of a time-reversal breaking excitonic condesate with a bulk spin spiral order. Through Hartree-Fock calculations we showed the difference between the time-reversal preserving and breaking excitonic condensates that can form in WTe$_2$ depending on the circumstances. Based on the mean field results, we performed quantum transport calculations for a finite width ribbon with disorder based on a tight-binding model. We demonstrated that while the time-reversal preserving excitonic insulator is topological and thus exhibits robust conductance quantization of edge state transport, the time-reversal breaking condensate deviates from $e^2/h$ per edge conductance in the presence of non-magnetic static disorder. Our results are in good agreement with the experimentally observed edge-length scaling of resistance with results close to quantized below 100 nm with the deviation increasing substantially for longer edges.

To provide some additional intuition for the mechanism behind the breakdown, we then supplemented these simulations by analytical edge transport calculations in the Luttinger liquid picture. Similarly to the previous work on the effect of RKKY nuclear spin spiral on helical edge states~\cite{Hsu2018}, we used the Schrieffer-Wolff transformation to capture the effective spin-spiral assisted back scattering from the non-magnetic impurities. This provides a qualitative understanding of the effective backscattering in disordered excitonic condensate with spin spiral order. Finally, by using bosonization, we determine the effect of this backscatering process on the transport properties of the system, determining the scaling dependence of resistance on length, temperature, and interaction strength. Our work thus provides a new mechanism which possibly contributes to the lack of perfect quantization in WTe$_2$ and paves the way for bridging the bulk and edge transport theory. Our results can also be generalized to transport in many other quantum physics systems with topologically protected edge states such as twisted multilayer graphene~\cite{Zhang2022local}.

\begin{acknowledgements}
This work was supported as part of the Center for Novel Pathways to Quantum Coherence in Materials, an Energy Frontier Research Center funded by the U.S. Department of Energy, Office of Science, Basic Energy Sciences (Y.-Q.W. and J.E.M.). M.P. was supported by the Quantum Science Center (QSC), a National Quantum Information Science Research Center of the U.S. Department of Energy (DOE).  M.P. received additional fellowship support from the Emergent Phenomena in Quantum Systems program of the Gordon and Betty Moore Foundation.
\end{acknowledgements}

\appendix

\noindent

\setcounter{section}{0}
\setcounter{equation}{0}
\setcounter{figure}{0}
\setcounter{table}{0}
\renewcommand\theequation{A\arabic{equation}}
\renewcommand\thefigure{A\arabic{figure}}
\renewcommand\thetable{A\arabic{table}}

\bibliography{apssamp}

\end{document}